\documentclass[10pt,a4paper,twoside]{article}
\usepackage{latexsym}
\begin{document}

\makeatletter
 \DeclareRobustCommand\cite{\unskip
   \@ifnextchar [{\@tempswatrue\@citex}{\@tempswafalse\@citex[]}}
 \def\@cite#1#2{$^{\hbox{\scriptsize{#1\if@tempswa , #2\fi}}}$}
\makeatother

\title{The Lie algebraic structure of extended Sutherland models}
\author{Kazuyuki Oshima\\
 \\
 Faculty of Education, Gifu Shotoku Gakuen University\\
 2078 Takakuwa, Yanaizu-cho, Hashima-gun\\
 Gifu 501-6194, Japan \\
 e-mail: kaz-oshima@muh.biglobe.ne.jp}

\date{\empty}

\maketitle
 
\begin{abstract}
 We disclose the Lie algebraic structure of two extended Sutherland models.
 Their Hamiltonians are $BC_{N}$, and $A_{N}$ Sutherland Hamiltonians with
 some additional terms.
 We show that both Hamiltonians can be written in the quadratic forms
 of generators of the Lie algebra $gl(N+1)$.

\bigskip


\medskip

\end{abstract}

\section{Introduction}
The quasi-exactly-solvable models have been developed considerably over
last decade. Ten years ago the crucial definition of quasi-exact-solvability
was made. \cite{Tur}
The author defined the quasi-exact-solvability as follows.
Let ${\cal P}_{n}$ be the space of polynomials of degree $\le n$.
\newtheorem{df}{Definition}
\begin{df}
Let us name a linear differential operator of the $k$-th order, $T_{k}$
{\em quasi-exactly-solvable}, if it preserves the space ${\cal P}_{n}$.
Correspondingly, the operator $E_{k}$, which preserves the infinite flag
$ {\cal P}_{0} \subset {\cal P}_{1} \subset {\cal P}_{2} \subset \cdots
\subset {\cal P}_{n} \subset \cdots $ of spaces of all polynomials, is
named {\em exactly-solvable}.
\end{df}
In particular, if operators can be represented by the elements of the
enveloping algebra of a certain Lie algebras realized in terms of differential 
operators preserving the space ${\cal P}_{n}$, they are quasi-exactly-solvable.
Following this concept, various quasi-exactly-solvable models have been proposed. 
In reference, \cite{MRT} the authors obtain the $sl_{2}$-based deformation of 
the Calogero models.
$A_{N},\,BC_{N},\,B_{N}$, $C_{N}$, and $D_{N}$ Calogero and Sutherland
Hamiltonian, as well as thier supersymmetric generalizations are shown to have
expression in the quadratic polynomials in the generators of the Lie algebra 
or the Lie superalgebra for the supersymmetric case. \cite{BTW}
Moreover the $sl(N+1)$ deformation of the Calogero models, whose Hamiltonians have 
the quadratic and sextic self-interaction terms, are shown to be 
quasi-exact-solvable. \cite{HS}  
Recently
the quasi-exact-solvable models are classified by using the new family of 
$A_{N}$-type Dunkl operators which includes the usual Dunkl operator. \cite{FGGRZ}

The purpose of this paper is to obtain the explicit Lie algebraic 
form of the extended Sutherland Hamiltonians, following the method
explained in the reference. \cite{BTW} 
We consider here two kinds of Hamiltonians.
One of them is an extended Hamiltonian of the $BC_{N}$-Sutherland model,
which has an additional term:
\begin{equation}
 \sum_{i=1}^{N} \left\{ g_{1} \cos^2 2x_{i} 
 + 2 g_{2} \cos 2x_{i}  \right\}.
\end{equation}
The other is an extended Hamiltonian of the $A_{N}$-Sutherland model,
which has an additional term:
\begin{equation}
 \sum_{i=1}^{N} \left\{ g_{1}\cos 4x_{i}+g_{2}\cos 2x_{i}+g_{3}\sin 4x_{i}
 +g_{4}\sin 2x_{i} \right\}.
\end{equation}
These Hamiltonians are indicated in the reference. \cite{FGGRZ}

This paper is arranged as follows.
In section 2, we consider an extended $BC_{N}$-Sutherland
Hamiltonian. First we rotate it with the ground state eigenfunction as a 
gauge fuctor. Then we rewrite the gauge rotated Hamiltonian in terms of
the elementary symmetric polynomials. Finally we arrive at the explicit
form of the Hamiltonian in quadratic form of generators which generate
the Lie algebra $gl(N+1)$. In section 3, we show an extended $A_{N}$-Sutherland
Hamiltonian has Lie algebraic form as the same technique as in section 2.

\section{Extended $BC_{N}$-Sutherland model}

In this section we begin with the case of the extended $BC_{N}$-Sutherland 
Hamiltonian.

\subsection{{\it Hamiltonian}}

The Hamiltonian for the Sutherland model of type $BC_{N}$ is \cite{BTW}
\begin{eqnarray}
H &=& -\sum_{i=1}^{N}\frac{\partial^2}{\partial x_{i}^2}+
      g \sum_{i \ne j} \left\{  \frac{1}{\sin^2(x_{i}-x_{j})}+
    \frac{1}{\sin^2(x_{i}+x_{j})} \right\}  \nonumber \\
    && \quad \quad \quad + 4 g_{3} \sum_{i=1}^{N}\frac{1}{\sin^2 2x_{i}} 
    + g_{4} \sum_{i=1}^{N} \frac{1}{\sin^2 x_{i}}
\label{BCH}
\end{eqnarray}
where $g,\,g_{3}$ and $g_{4}$ are coupling constants. 
From the Hamiltonian (\ref{BCH}) we can deduce the Hamiltonian of the
Sutherland models of 
type $B_{N},\,C_{N},\,D_{N}$ as follows:

\medskip

\begin{tabular}{ccc}
  $g_{3}=0$ &:& $B_{N}$, \\
  $g_{4}=0$ &:& $C_{N}$, \\
  $g_{3}=g_{4}=0$ &:& $D_{N}$.
\end{tabular}

\medskip

We will consider a Hamiltonian with the external potential \cite{FGGRZ}
\begin{equation}
 \sum_{i=1}^{N} \left\{ g_{1} \cos^2 2x_{i} 
 + 2 g_{2} \cos 2x_{i} \right\},
\end{equation}
that is, 
\begin{eqnarray}
H&=&-\sum_{i=1}^{N}\frac{\partial^2}{\partial x_{i}^2}
       + g \sum_{i \ne j} \left\{  \frac{1}{\sin^2(x_{i}-x_{j})}+ \label{extH}
    \frac{1}{\sin^2(x_{i}+x_{j})} \right\} \\\nonumber
    &&+\sum_{i=1}^{N}\left\{g_{1} \cos^2 2x_{i} +2 g_{2} \cos 2x_{i} +
     4 g_{3} \frac{1}{\sin^2 2x_{i}} +g_{4}\frac{1}{\sin^2 x_{i}} \right\} \\\nonumber
\end{eqnarray}
where 
\begin{eqnarray*}
 g &=& \nu(\nu-1), \\
 g_{1} &=& -\nu_{1}^2, \\
 g_{2} &=&  \nu_{1} \left\{1+2n + 2(N-1)\nu+\nu_{3}+\nu_{4} \right\} 
            \quad (n \in {\bf Z}),\\
 g_{3} &=&  -\nu_{3}(\nu_{3}-1), \\
 g_{4} &=&  \nu_{3}(\nu_{3}-1)-\nu_{4}(\nu_{4}-1).
\end{eqnarray*}
Let us call the Hamiltonian (\ref{extH}) the extended $BC_{N}$-Sutherland Hamiltonian.

\subsection{{\it Gauge rotation}}

First we make a gauge rotation of the Hamiltonian (\ref{extH}) with the ground state
eigenfunction
\begin{eqnarray}
 \mu &=& \left\{\prod_{i<j} \sin(x_{i}-x_{j}) \sin(x_{i}+x_{j})\right\}^{\nu} \\\nonumber
      && \times \left\{\prod_{i=1}^{N}\cos x_{i} \right\}^{\nu_{3}}
         \left\{\prod_{i=1}^{N}\sin x_{i} \right\}^{\nu_{4}}
         \prod_{i=1}^{N}\exp \left(-\frac{\nu_{1}}{2}\cos 2x_{i} \right)
\end{eqnarray}
as gauge factor, $\widetilde{H}=\mu^{-1} H \mu $.
Using the relation
\begin{eqnarray}
  && \sum_{i=1}^{N} \left[\sum_{j(\ne i)} 
  \left( \frac{ \cos(x_{i}-x_{j})}{\sin(x_{i}-x_{j})} +
         \frac{ \cos(x_{i}+x_{j})}{\sin(x_{i}+x_{j})} \right)\right]^2 \\\nonumber
  &&= 2 \sum_{i < j} \left( \frac{ \cos^{2}(x_{i}-x_{j})}{\sin^{2}(x_{i}-x_{j})} +
                 \frac{ \cos^{2}(x_{i}+x_{j})}{\sin^{2}(x_{i}+x_{j})} \right)+
 constant, \\ \nonumber
\end{eqnarray}
it is not difficult to obtain that
\begin{eqnarray}
 \widetilde{H} &=& 
  -\sum_{i=1}^{N} \frac{\partial^2}{\partial x_{i}^2}
    -2 \sum_{i=1}^{N} \Bigg\{\nu 
    \sum_{j (\ne i)}\left(\frac{\cos (x_{i}-x_{j})}{\sin (x_{i}-x_{j})}+
     \frac{\cos (x_{i}+x_{j})}{\sin (x_{i}+x_{j})} \right) \label{rotH} \\\nonumber
   && \quad \quad \quad + (\nu_{3}+\nu_{4}) \frac{\cos 2 x_{i}}{\sin 2 x_{i}}  
    -(\nu_{3}-\nu_{4})\frac{1}{\sin 2x_{i}}
    +\nu_{1} \sin 2x_{i} \Bigg\}\frac{\partial}{\partial x_{i}} \\\nonumber
    && \quad \quad \quad +4n \nu_{1} \sum_{i=1}^{N} \cos 2x_{i},
\end{eqnarray}
hereafter we will omit the constant terms.

\subsection{{\it Lie algebraic form}}

In order to disclose the Lie algebraic structure of the gauge rotated 
Hamiltonian (\ref{rotH}), it is expedient to rewrite 
the Hamiltonian ${\widetilde H}$ in terms of 
the elementary symmetric polynomials.
Let $e_{i}$ be the $i$-th elementary symmetric polynomials of $\cos 2x$,
\begin{eqnarray*}
 e_{1} &=& \sum_{i=1}^{N}\cos 2x_{i}, \\
 e_{2} &=& \sum_{i <j} \cos 2x_{i} \cos 2x_{j}, \\
 e_{3} &=& \sum_{i<j<k} \cos 2 x_{i} \cos 2 x_{j} \cos 2 x_{k}, \\
 \vdots     && \\
 e_{N} &=& \cos 2x_{1} \cos 2x_{2} \cdots \cos 2x_{N}.
\end{eqnarray*}

\newtheorem{lemma}{Lemma}
\begin{lemma}
In these variables $\{e_{1},e_{2},\dots,e_{N}\}$, the gauge rotated 
Hamiltonian $\widetilde{H}$ becomes
\begin{eqnarray}
 \widetilde{H} &=& -4 \sum_{k,l=1}^{N}\Bigg\{ N e_{k-1} e_{l-1}-
 \sum_{i \ge 0} \bigg[ (k-i)e_{k-i}e_{l+i} 
 +(l-1+i)e_{k-1-i}e_{l-1+i} \nonumber\\
 && \quad \quad -(k-2-i)e_{k-2-i}e_{l+i} 
        -(l+1+i)e_{k-1-i}e_{l+1+i}\bigg] \Bigg\}
    \frac{\partial}{\partial e_{k}} \frac{\partial}{\partial e_{l}} \nonumber\\
 && +4 \sum_{k=1}^{N} \Bigg\{ \nu(N-k+2)(N-k+1)e_{k-2} \nonumber\\
 && \quad \quad \quad +\bigg[ \nu k(2N-k-1)+2\nu_{3}k +k \bigg]e_{k} \nonumber\\
 && \quad \quad \quad +\nu_{1}\bigg[-e_{1}e_{k}+(N-k+1)e_{k-1}
      +(k+1)e_{k+1} \bigg] \Bigg\}
    \frac{\partial}{\partial e_{k}} \nonumber\\
 && +4n \nu_{1} e_{1}.
\label{covH}
\end{eqnarray} 
\end{lemma}

\noindent{\bf Proof.} \enskip
After the change of variables $x_{i} \to e_{i}$, the Hamiltonian (\ref{rotH}) is
\begin{eqnarray}
 \widetilde{H} &=& 
  - \sum_{k,l=1}^{N} \sum_{i=1}^{N} \frac{\partial e_{k}}{\partial x_{i}}
     \frac{\partial e_{l}}{\partial x_{i}}\frac{\partial}{\partial e_{k}}
     \frac{\partial}{\partial e_{l}} 
  - \sum_{k=1}^{N} \sum_{i=1}^{N} \frac{\partial^{2} e_{k}}{\partial x_{i}^{2}}
     \frac{\partial}{\partial e_{k}} \\\nonumber
    && \quad - \nu \sum_{k=1}^{N} \sum_{i \ne j} \Bigg\{ 
      \frac{\cos (x_{i}-x_{j})}{\sin (x_{i}-x_{j})}
      \left( \frac{\partial e_{k}}{\partial x_{i}} 
     -\frac{\partial e_{k}}{\partial x_{j}} \right) \\\nonumber
     && \quad \quad \quad  \quad  \quad \quad + \frac{\cos (x_{i}+x_{j})}{\sin (x_{i}+x_{j})} 
      \left( \frac{\partial e_{k}}{\partial x_{i}}
       +\frac{\partial e_{k}}{\partial x_{j}} \right) \Bigg\} 
        \frac{\partial}{\partial e_{k}} \\\nonumber
   && \quad \quad \quad 
     -2 (\nu_{3}+\nu_{4}) \sum_{k=1}^{N}\sum_{i=1}^{N} \frac{\cos 2 x_{i}}{\sin 2 x_{i}} 
        \frac{\partial e_{k}}{\partial x_{i}} \frac{\partial}{\partial e_{k}} \\\nonumber
    && \quad \quad \quad
     +2 (\nu_{3}-\nu_{4})\sum_{k=1}^{N}\sum_{i=1}^{N}  \frac{1}{\sin 2x_{i}}
        \frac{\partial e_{k}}{\partial x_{i}}\frac{\partial}{\partial e_{k}} \\\nonumber
      && \quad \quad \quad -2 \nu_{1} \sum_{k=1}^{N}\sum_{i=1}^{N}  
        \sin 2x_{i} \frac{\partial e_{k}}{\partial x_{i}}
        \frac{\partial}{\partial e_{k}}+4n \nu_{1} \sum_{i=1}^{N} \cos 2x_{i}.
\end{eqnarray}
As an example, let us demonstrate how to express
\begin{equation}
 X_{k}=\sum_{i=1}^{N} \sin 2x_{i} \frac{\partial e_{k}}{\partial x_{i}} 
\end{equation}
in terms of the elementary symmetric polynomials. Let us consider
the generating function of $X_{k}$:
\[
X(t)=\sum_{k=0}^{N}X_{k}t^{k}.
\]
Note that the generating function of elementary symmetric polynomials $e_{k}$ is
\begin{equation}
 E(t)=\sum_{k=0}^{N}e_{k}t^{k}
 =\exp\left[ \sum_{n=1}^{\infty} \frac{(-1)^{n+1}}{n}p_{n}(\cos 2x)t^{n} \right],
\label{genE}
\end{equation}
where $p_{n}$ are the power sums:
\[
 p_{n}(\cos 2x)=\sum_{i=1}^{N} \cos^{n} 2x_{i}.
\]
Calculating the generating function $X(t)$, we have
\begin{eqnarray}
 X(t) &=& \sum_{k=0}^{N} \left[ \sum_{i=1}^{N}\sin 2x_{i} 
   \frac{\partial e_{k}}{\partial x_{i}} \right]t^{k} \\\nonumber
  &=& \sum_{i=1}^{N}\sin 2 x_{i} \sum_{k=0}^{N} 
        \frac{\partial e_{k}}{\partial x_{i}} t^{k} \\\nonumber
  &=& \sum_{i=1}^{N} \sin 2 x_{i} \frac{\partial E(t)}{\partial x_{i}} \\\nonumber
  &=& 2 \left\{\sum_{n=1}^{\infty}(-1)^{n}p_{n-1}(\cos 2x)t^{n}-
         \sum_{n=1}^{\infty}(-1)^{n}p_{n+1}(\cos 2x)t^{n} \right\}E(t).
\end{eqnarray}
It is not difficult to show that
\begin{equation}
X(t)=2\left(p_{1}(\cos 2x)-Nt+t^{2}\frac{\partial}{\partial t}-
       \frac{\partial}{\partial t} \right) E(t).
\label{diffE}
\end{equation}
Substituting $E(t)=\sum_{k=0}^{N}e_{k}t^{k}$ in (\ref{diffE}), we find that
\begin{equation}
 X_{k}=2 \left\{e_{1}e_{k}-(N-k+1)e_{k-1}-(k+1)e_{k+1} \right\}.
\end{equation}

One can show the other parts similarly.
\hfill $\Box$

\medskip

Next we will realize the Lie algebra $gl(N+1)$ in terms of elementary symmetric
polynomials. One of the simplest realization of $gl(N+1)$ is the differential
realization.
The generators can be represented in the following form:
\begin{eqnarray}
 J_{i}^{-} &=& \frac{\partial}{\partial e_{i}} \quad \quad
  (i=1,2,\dots,N), \label{jm} \\
 J_{ij}^{0} &=& e_{i} \frac{\partial}{\partial e_{j}} \quad
  (i,j=1,2,\dots,N), \\
 J^{0} &=& n -\sum_{i=1}^{N} e_{i}\frac{\partial}{\partial e_{i}},  \\
 J_{i}^{+} &=& e_{i}J^{0} \quad (i=1,2,\dots,N), \label{jp}
\end{eqnarray}
where the parameter $n$ is an integer.

If $n$ is a non-negative integer, the generators
act on the representation space of 
polynomials in $N$ variables of the following type
\begin{equation}
 {\cal P}_{n} = {\rm span}\left\{e_{1}^{n_{1}}e_{2}^{n_{2}}\cdots e_{N}^{n_{N}} \, \Biggm|
 \, 0 \le \sum_{i=1}^{N} n_{i} \le n \right\}.
\end{equation}

With simple algebraic transformations from (\ref{covH}), 
we obtain the following result.

\newtheorem{prop}{Proposition}
\begin{prop}
We obtain the Lie algebraic form of the gauge rotated extended 
$BC_{N}$-Sutherland Hamiltonians (\ref{covH}) as follows:
\begin{eqnarray}
 && \widetilde{H} = \nonumber\\
 && -4\sum_{k,l=1}^{N}\Bigg\{ N J_{k-1,l}^{0}J_{l-1,k}^{0}-
 \sum_{i \ge 0} \bigg[
 (k-i)\big(J_{k-i,l}^{0}J_{l+i,k}^{0}-\delta_{i0}J_{k,k}^{0}\big) \nonumber\\
 && \quad  +(l-1+i)\big(J_{k-1-i,l}^{0}J_{l-1+i,k}^{0}
       -\delta_{i1}J_{k-2,k}^{0}\big) \nonumber\\
 && \quad  -(k-2-i)\big(J_{k-2-i,l}^{0}J_{l+i,k}^{0}-\delta_{i0}J_{k-2,k}^{0}\big)
    -(l+1+i)J_{k-1-i,l}^{0}J_{l+1+i,k}^{0} \bigg] \Bigg\} \nonumber\\
 && +4\sum_{k=1}^{N}\Bigg\{\nu(N-k+2)(N-k+1)J_{k-2,k}^{0} 
    +\nu_{1}(N-k+1)J_{k-1,k}^{0} \nonumber\\
 && \quad  +\bigg[\nu k (2N-k-1)+2 \nu_{3} k +k \bigg]J_{k,k}^{0}
    +\nu_{1}(k+1)J_{k+1,k}^{0} \Bigg\}  \nonumber \\
 && +4\nu_{1}J_{1}^{+}.
\end{eqnarray}
Here we define
$$ J_{0i}^{0}=J_{i}^{-} \quad (i=1,2,\dots,N). $$
\end{prop}
 
\section{Extended $A_{N}$-Sutherland Hamiltonian}

In this section we devote to a consideration of the extended $A_{N}$-Sutherland 
Hamiltonian with using the same approach as in the
previous section.

\subsection{{\it Hamiltonian}}

A second Hamiltonian we deal with is an extension of $A_{N}$-Sutherland model:
\begin{equation}
  H = -\sum_{i=1}^{N} \frac{\partial^2}{\partial x_{i}^2} 
        + g \sum_{i \ne j} \frac{1}{\sin^2(x_{i}-x_{j})}. 
\end{equation}
Considering the external potential \cite{FGGRZ}
\begin{equation}
 2 \sum_{i=1}^{N} \left\{ g_{1}\cos 4x_{i}+2 g_{2}\cos 2x_{i}+2 g_{3}\sin 4x_{i}
 +2 g_{4}\sin 2x_{i} \right\},
\end{equation}
we define a Hamiltonian
\begin{eqnarray}
 H &=& -\sum_{i=1}^{N} \frac{\partial^2}{\partial x_{i}^2} \label{extH2}
        + g \sum_{i \ne j} \frac{1}{\sin^2(x_{i}-x_{j})} \nonumber\\
  && +2 \sum_{i=1}^{N} \left\{ g_{1} \cos 4x_{i}+2 g_{2} \cos 2 x_{i}
        +2 g_{3} \sin 4x_{i} + 2 g_{4} \sin 2x_{i} \right\} 
\end{eqnarray}
where coupling constants $g, g_{1}, g_{2}, g_{3}, g_{4}$ satisfy
\begin{eqnarray*}
 g &=& \nu(\nu-1), \\
 g_{1} &=& (\beta-\gamma)(\beta+\gamma), \\
 g_{2} &=& (\gamma+\alpha \beta+ n \gamma +\nu \gamma(N-1)), \\
 g_{3} &=&  \beta \gamma, \\
 g_{4} &=& (\alpha \gamma -\beta- n \beta -\nu \beta(N-1)).
\end{eqnarray*}
We call the Hamiltonian (\ref{extH2}) the extended $A_{N}$-Sutherland Hamiltonian.

\subsection{{\it Gauge rotation}}
Similarly to what was done in section 2 for the extended $BC_{N}$-Sutherland Hamiltonian,
we first make a gauge rotation of (\ref{extH2}) with the ground state eigenfunction
\begin{equation}
\mu(x) =\prod_{i<j} \sin^{\nu}(x_{i}-x_{j})\prod_{i=1}^{N}\cos^{n}x_{i}
        \exp \left[\alpha x_{i}+\beta \sin2x_{i}-\gamma \cos 2x_{i} \right]
\end{equation}
as a gauge fuctor ${\widetilde H}=\mu^{-1}H\mu$. It is not difficult to obtain
\begin{eqnarray}
 {\widetilde H} &=&  -\sum_{i=1}^{N}\frac{\partial^2}{\partial x_{i}^2} 
      -2 \sum_{i=1}^{N} \Bigg\{\nu \sum_{i \ne j}
      \frac{\cos(x_{i}-x_{j})}{\sin(x_{i}-x_{j})}-n \frac{\sin x_{i}}{\cos x_{i}}  \label{rotH2}
     \nonumber \\
  && \quad \quad \quad
      +(\alpha+2 \beta \cos 2x_{i} +2 \gamma \sin 2x_{i}) \Bigg\}
      \frac{\partial}{\partial x_{i}} \\
  && \quad - n(n-1)\sum_{i=1}^{N} \tan^2 x_{i} 
    + 2 \nu n \sum_{i <j} \tan x_{i} \tan x_{j} \nonumber \\
   && \quad \quad + 2 n (\alpha -2 \beta) \sum_{i=1}^{N} \tan x_{i}. \nonumber
\end{eqnarray}

\subsection{{\it Lie algebraic form}}

Next we rewrite the gauge rotated Hamiltonian (\ref{rotH2}) in terms of the 
elementary symmetric polynomials of $\tan x$.
\begin{eqnarray*}
 e_{1} &=& \sum_{i=1}^{N} \tan x_{i}, \\
 e_{2} &=& \sum_{i<j} \tan x_{i} \tan x_{j}, \\
 e_{3} &=& \sum_{i<j<k} \tan x_{i} \tan x_{j} \tan x_{k}, \\
       &\vdots& \\
 e_{N} &=& \tan x_{1} \tan x_{2} \cdots \tan x_{N}.
\end{eqnarray*}

Numerous but straightforward calculations similar to what we made 
in the proof of lemma 1 leads to the following representation.

\begin{eqnarray}
 {\widetilde H} &=& -\sum_{k,l=1}^{N}\Bigg\{N e_{k-1}e_{l-1}-e_{1}e_{k+1}e_{l}-
 e_{1}e_{k}e_{l+1}+(e_{1}^{2}-2e_{2})e_{k}e_{l} \label{covH2} \nonumber \\
 && +\sum_{i \ge 0}\bigg[(k-2-i)e_{k-2-i}e_{l+i}-(l-1+i)e_{k-1-i}e_{l-1+i} \nonumber \\
  &&  \quad \quad +2(k-i)e_{k-i}e_{l+i}-2(l+1+i)e_{k-1-i}e_{l+1+i} \nonumber \\
  &&  \quad \quad +(k+2-i)e_{k+2-i}e_{l+i}-(l+3+i)e_{k-1-i}e_{l+3+i} \bigg] \Bigg\}
      \frac{\partial}{\partial e_{k}} \frac{\partial}{\partial e_{l}}  \nonumber \\
  && + \nu \sum_{k=1}^{N} \Bigg\{(N-k+2)(N-k+1)e_{k-2} \nonumber \\
    && \quad \quad    -2 \bigg[ k(N-k)+e_{2} \bigg]e_{k} +(k+1)k e_{k+2} \Bigg\} 
       \frac{\partial}{\partial e_{k}} \nonumber \\
  && -\sum_{k=1}^{N}\Bigg\{2\bigg[(k+2)e_{k+2}-e_{1}e_{k+1}+(k+e_{1}^2-2e_{2})e_{k}\bigg]
      \Bigg\} \frac{\partial}{\partial e_{k}}  \nonumber \\
  && +2n \sum_{k=1}^{N}\Bigg\{ (k+2)e_{k+2}-e_{1}e_{k+1}+(k+e_{1}^{2}-2e_{2})e_{k}\Bigg\}
      \frac{\partial}{\partial e_{k}} \nonumber \\
  && -2\alpha \sum_{k=1}^{N} \Bigg\{\bigg[N-(k-1)\bigg]e_{k-1}+e_{1}e_{k}-(k+1)e_{k+1} \Bigg\}
      \frac{\partial}{\partial e_{k}}  \nonumber \\
  && -4 \beta \sum_{k=1}^{N} \Bigg\{\bigg[N-(k-1)\bigg]e_{k-1}-e_{1}e_{k}+(k+1)e_{k+1} \Bigg\}
      \frac{\partial}{\partial e_{k}}  \nonumber \\
  && -4 \gamma \sum_{k=1}^{N}2k e_{k} \frac{\partial}{\partial e_{k}}   \nonumber \\
  && -n(n-1)(e_{1}^{2}-2e_{2})+2\nu n e_{2}+2 n(\alpha-2 \beta)e_{1}.
\end{eqnarray}

Now we use the same realization (\ref{jm})-(\ref{jp}) of Lie algebra $gl(N+1)$. With 
simple algebraic transformations from (\ref{covH2}), we derive the following result.

\begin{prop}
We obtain the Lie algebraic form of the gauge rotated extended $A_{N}$-Sutherland Hamiltonian
(\ref{covH2}) as follows:
\begin{eqnarray}
 &&{\widetilde H} = \nonumber \\ 
    && -\sum_{k,l=1}^{N}\Bigg\{ N J_{k-1,l}^{0}J_{l-1,k}^{0} 
   +\sum_{i \ge 0}\bigg[ (k-2-i)(J_{k-2-i,l}^{0}J_{l+i,k}^{0}-\delta_{i0}J_{k-2,k}^{0}) \nonumber \\
  && \quad  -(l-1+i)(J_{k-1-i,l}^{0}J_{l-1+i,k}^{0}-\delta_{i1}J_{k-2,k}^{0}) \nonumber \\
  && \quad  +2(k-i)(J_{k-i,l}^{0}J_{l+i,k}^{0}-\delta_{i0}J_{k,k}^{0}) 
     -2(l+1+i) J_{k-1-i,l}^{0}J_{l+1+i,k}^{0} \nonumber \\
  && \quad  +(k+2-i)(J_{k+2-i,l}^{0}J_{l+i,k}^{0}-\delta_{i0}J_{k+2,k}^{0}) 
     -(l+3+i) J_{k-1-i,l}^{0}J_{l+3+i,k}^{0} \bigg] \Bigg\} \nonumber \\
  && +\sum_{k=1}^{N} \Bigg\{ \nu (N-k+1)(N-k+2)J_{k-2,k}^{0} 
     -\bigg[2k \nu (N-k)+8 \gamma k \bigg]J_{k,k}^{0} \nonumber \\
  && \quad \quad -2(\alpha+2 \beta)(N-k+1)J_{k-1,k}^{0} \nonumber \\
  && \quad \quad  +\bigg[\nu k(k+1)+2(n-1)(k+2)\bigg]J_{k+2,k}^{0} \Bigg\} \\
  && -2 \sum_{k=1}^{N}J_{k+1,k}^{0}J_{1}^{+}+2(n+\nu+1)J_{2}^{+}+2(\alpha-2 \beta)J_{1}^{+}
     -2 \sum_{k=1}^{N}J_{k,k}^{0}J_{2}^{+}-J_{1}^{+}J_{1}^{+}. \nonumber
\end{eqnarray}
Here we define $J_{0i}^{0}=J_{i}^{-}$ (i=1,2,\dots,N).
\end{prop}

\section*{Acknowledgements}

I am thankful to Professor Hidetoshi Awata for his continuous
encouragement and helpful comments.

\bigskip

{\it Note added.}
After submission of this paper, an article \cite{Tan} has been brought to the
my attention.
In that article, the complete classification of the quasi-exactly solvable models from the
sl(M+1) generators are given.
I am thankful to Professor Toshiaki Tanaka for informing his articles.


\begin{thebibliography}{9}
\makeatletter
 \def\@biblabel#1{{#1.}}
\makeatother
 
 \bibitem{Tur} A. Turbiner,
   {\it Contemp.Math.} {\bf Vol.160} 263-310 (1994).
 
 \bibitem{MRT} A. Minzoni, M. Rosenbaum and A. Turbiner,
   {\it Mod.Phys.Lett.} {\bf A11} 1977-1984 (1996).
 
 \bibitem{BTW} L. Brink, A. Turbiner and N. Wyllard,
   {\it J.Math.Phys.} {\bf 39} 1285-1315 (1998).

 \bibitem{HS} Xinrui Hou and M. Shifman,
   {\it Int.J.Mod.Phys.} {\bf A14} 2993-3004 (1999).

 \bibitem{FGGRZ} F. Finkel, D. G\'{o}mez-Ullate, A. Gonz\'{a}lez-Lop\'{e}z,
  M. A. Rodr\'{\i}guez 
  and R. Zhdanov,
   {\it Comm.Math.Phys.} {\bf 221} 472-496 (2001).

 \bibitem{Tan} Toshiaki Tanaka,
   {\it Annals Phys.} {\bf 309} 239-280 (2004).
\end{thebibliography}
\end{document}